\documentclass[12pt]{iopart}

\usepackage{iopams}  
\usepackage{graphicx} 
\usepackage{bm}
\begin{document}

\title{D-brane solitons and boojums in field theory and Bose-Einstein condensates}

\author{Kenichi Kasamatsu$^1$, Hiromitsu Takeuchi$^2$, and Muneto Nitta$^3$}

\address{$^1$ Department of Physics, Kinki University, Higashi-Osaka, 577-8502, Japan}
\address{$^2$ Graduate School of Integrated Arts and Sciences, Hiroshima
University, Kagamiyama 1-7-1, Higashi-Hiroshima 739-8521, Japan}
\address{$^3$ Department of Physics, and Research and Education Center for Natural 
Sciences, Keio University, Hiyoshi 4-1-1, Yokohama, Kanagawa 223-8521, Japan}
\ead{kenichi@phys.kindai.ac.jp}
\begin{abstract}

In certain field theoretical models, composite solitons
consisting of a domain wall and vortex lines attached to the wall have
been referred to as D-brane solitons.
We show that similar composite solitons can be realized in 
phase-separated two-component Bose-Einstein condensates.
We discuss the similarities and differences between topological solitons in the Abelian-Higgs
model and those in two-component Bose-Einstein condensates.
Based on the formulation of gauge theory,
we introduce the ``boojum charge'' to characterize the ``D-brane soliton'' in
Bose-Einstein condensates.

\end{abstract}

\maketitle

\section{Introduction}\label{Intro}
String theory is the most promising candidate for producing
a unified theory of the four fundamental forces of nature
\cite{Polchinskibook}.
Before the so-called second revolution of string theory, five kinds of
theories (typeI/IIA/IIB/heterotic SO/heterotic E8) were
known to be consistent string theories, which were formulated
in perturbation in terms of a genus expansion
of the world-sheet \cite{Johnson:2003gi}. The discovery of the Dirichlet branes 
(D-branes) was a turning point for string theorists, showing 
that all perturbative string theories are related
non-perturbatively. D-branes have been found to be
non-perturbative solitonic states of string theory and are 
characterized as hypersurfaces on which open fundamental
strings can terminate with the Dirichlet boundary
condition. D-branes are the most fundamental
tool for studying the non-perturbative dynamics of string
theory, which would enable the solving of the problems such
as the dimensionality of the universe and the generation
number of quarks and leptons

In some field theories, topological solitons consisting of a complex of domain walls 
and strings have been proposed as a prototype of the original D-brane in string
theory, referred to as the ``D-brane soliton". The D-brane soliton was first discussed by Gauntlett
{\it et al}. based on the hyper-K\"{a}hler nonlinear sigma model (NL$\sigma$M) \cite{Gauntlett}. 
A similar argument was extended to Abelian or non-Abelian gauge theory with 
finite gauge coupling \cite{Shifman1,Shifman2,Isozumi,Eto}. 
All possible composite solitons of domain walls, vortices, monopoles, and instantons, including
D-brane solitons in field theoretical models are reviewed in Ref. \cite{Eto:2006pg}. 
The correspondence between the D-brane in string theory and
that in field theory was established by the following facts \cite{Tong}: (i) The branes should have a localized U(1)
Nambu-Goldstone mode, which can be rewritten as the U(1)
gauge field on its world volume. The effective description
of the collective motion of the D-branes is then 
given by the Dirac-Born-Infeld (DBI) action in the
low energy regime \cite{Leigh}. The DBI action is a nonlinear 
action of the scalar field (corresponding to the transverse position
of the D-brane) and the U(1) gauge field \cite{Dirac}.  
(ii) Vortex lines attached to the
domain wall are analogous to fundamental strings because
their endpoints are electrically charged and identical
to solitons known as ``BIons" in the DBI action \cite{Gibbons:1997xz}.

Based on these findings, laboratory experiments of D-brane physics have been proposed that 
exploit the fact that some aspects of these field theoretical models are closely related to the theory of 
condensed matter phenomena, and thus there are several systems appearing in nature 
which can be said to admit analogs of D-branes. 
The purpose of the analogy argument is to find a way to observe at least some parts of the original 
complex theory in cosmology or high energy physics by simulating the analogous phenomena 
in condensed matter experiments. 
To date, D-branes have only been a theoretical hypothesis and 
it is not clear whether or not they can be used to describe nature or not. 
Therefore, it is important to simulate D-brane physics 
in a laboratory to raise the practical applicability of this hypothesis 
and to further develop D-brane physics. 
For example, analogues of ``branes" have been 
discussed in an experiment on superfluid $^{3}$He \cite{Bradley}, where 
Brane--anti-brane collision was simulated using the boundary between $^3$He-A and $^3$He-B phases 
in a well controlled manner; the brane--anti-brane collision could explain the Big Bang in brane cosmology, 
after which lower-dimensional branes, e.g., cosmic strings, remain as relics. 
However, their exact correspondence to D-branes in string theory was not clarified in Ref. \cite{Bradley}. 
In other systems, e.g., ferromagnets, it is difficult to create and observe three-dimensional 
topological defects experimentally, so this kind of study has not yet been reported. 

Recently, we reported that an analogue of a D-brane can be realized 
in ultra-cold atomic Bose-Einstein condensates (BECs) and the experimental verification 
of various D-brane dynamics could be possible \cite{KasamatsuD,Takeuchitac,Nittavorton}. 
The domain wall in two-component BECs can mimic the fundamental properties of the D-brane, 
because the theoretical formulation in two-component BEC can be mapped to 
the NL$\sigma$M by introducing a pseudospin representation of the order 
parameter \cite{Babaev,Kasamatsu2,KTUreview}. 
It can then be shown that the domain wall in a two-component BEC has 
a localized U(1) Nambu-Goldstone mode, which is related to the 
local U(1) gauge field on the wall as a necessary degree of freedom 
for the effective description with the DBI action of 
a D-brane \cite{Leigh,Dirac,Gibbons:1997xz}. Thus, the resultant wall-vortex composite 
solitons could possess the characters (i) and (ii) described above and correspond to the 
D-branes in the sense of Ref. \cite{Gauntlett}. 
Here, the domain wall and the vortex can be identified as the D-brane and the 
fundamental string, respectively. 
Compared to the above systems, two-component BECs have a great advantage for 
exploring D-brane physics, because several experimental techniques to create, control, and 
observe the topological defects have been very well established \cite{nonlinearbook}. 
 
In this paper, we discuss the correspondence of topological solitons between BEC systems 
and gauge theoretical models to clarify the similarities and differences. 
We study various topological solitons such as vortices, domain walls, and their complexes, 
i.e., D-brane solitons, in the Abelian--multi-Higgs model, which is a similar formulation to 
the two-component Gross-Pitaevskii (GP) model that governs the dynamics of BECs 
within the mean-field level. The GP model corresponds to the strong gauge coupling limit 
of the Abelian-Higgs model, where the gauge field is not an independent dynamical degree 
of freedom and is consistently determined by the configuration of the matter (Higgs) fields. 
We also focus on the topological structure of the wall-vortex connecting points in D-brane solitons. 
In a field theoretical model, these points form defects called ``boojum" as the negative binding
energy of vortices and a wall, and a half of the negative
charge of a single monopole \cite{Isozumi,Sakai,Auzzi}. 
Boojums are known as point defects existing upon the surface of the ordered phase; the name was 
first introduced to physics by Mermin 
in the context of superfluid $^3$He \cite{Mermin}. 
Boojums can exist in different physical systems, such as the interface separating A and B phases 
of superfluid $^3$He \cite{Blaauwgeers,Volovik}, 
liquid crystals \cite{Kleman}, the Langmuir monolayers at air-water interfaces \cite{Fischer}, 
multi-component BECs with a spatially tuned interspecies interaction \cite{Takeuchi,Borgh}, 
and high density quark matter \cite{Cipriani}.
In the present model, boojums can be found at the end points of vortices on the domain wall, at which 
the vortices change their character from singular to coreless type. 

This paper is organized as follows. In Sec. \ref{Dbranefield}, 
we first review briefly topological solitons in the Abelian-Higgs model.  
The analysis of this section can be applied directly to two-component 
BECs. In Sec. \ref{formulation}, we discuss the formulation of two-component BECs 
and discuss the similarities and differences by comparing the structure of topological 
solitons such as vortices and D-brane solitons of the field theoretical 
model in Sec. \ref{Dbranefield}.
We conclude this paper in Sec. \ref{concle}. 

\section{D-brane solitons in field theoretical models}\label{Dbranefield}
As described in Sec. \ref{Intro}, similar structures of D-branes may occur in various field theories. 
In a field theory, both branes and strings arise as solitonic type 
objects. In this section, we study prototypes of topological solitons including vortices, domain walls, and
D-brane solitons in gauge theoretical models. The analog of a D-brane in the NL$\sigma$M was first 
pointed out by Gauntlett {\it et al.}, where they referred to it as a Q-kink-lump solution \cite{Gauntlett}. 
It turns out that the low energy effective theory of the collective coordinate for this solution 
can be described by the DBI action \cite{Leigh}. 
The vortices attached to the wall are identified as 
fundamental strings since their endpoints are electrically charged 
and identical to BIons in the DBI action \cite{Gibbons:1997xz}.
Therefore, the NL$\sigma$M offers a simplified model for studying 
D-brane dynamics, that is instructive for studying full string theory. 
A similar discussion has subsequently been made for supersymmetric gauge theory \cite{Shifman1}. 
For the strong coupling limit, the model is reduced to the NL$\sigma$M and the 
analytical solutions of the D-brane soliton can 
be obtained \cite{Isozumi}. 
This gauge theoretical model is closely related with the 
two-component GP model, 
where two-component scalar fields 
(order parameters) represent the condensate wave function of the BECs. 
The analysis of this section can be directly applied to our problem in some limit. 
Importantly, from this formulation, we can define a suitable topological charge that 
characterizes the D-brane soliton from 
which we can understand the details of the connecting 
points, i.e., boojum, of the wall and the vortex. 

\subsection{Vortex in Abelian-Higgs model}\label{AbeHigtheory}
We start from the simplest Abelian-Higgs model to understand the structure of the 
topological defects in the field theoretical model, which is useful in the following discussion. 
The Abelian-Higgs model is given by 
\begin{eqnarray}
H = \int d^3 x \left[ \frac{1}{2e^2} ({\bf E}^2 + {\bf B}^2) + |(\nabla - i {\bf A}) \phi|^2 + \frac{\lambda}{4} (|\phi|^2 - v^2)^2 \right], 
\end{eqnarray}
where $e$ is the gauge coupling (electric charge), 
$\lambda$ the Higgs scalar coupling and $v$ 
the vacuum expectation value (VEV) of the scalar field $\phi$, and 
${\bf E}$ and ${\bf B}$ are the electronic and magnetic field, respectively. 
For the vector potential ${\bf A}$, we have 
taken the transformation from the conventional notation as $e {\bf A} \to {\bf A}$ for convenience. 
The model is identical to the Ginzburg-Landau free energy in the theory of type-II superconductor. 
The hamiltonian is invariant under the local U(1) transformation 
$\phi({\bf r}) \to e^{i \alpha({\bf r})} \phi({\bf r})$ and ${\bf A} \to {\bf A} + \nabla \alpha({\bf r})$. 
The stationary point of this hamiltonian gives the equations 
\begin{eqnarray}
(\nabla - i  {\bf A})^2 \phi = \frac{\lambda}{2} (|\phi|^2 - v^2) \phi, \label{abehigseq} \\
{\bf j} = ie^2 \left[ \phi^{\ast} \nabla \phi - (\nabla \phi^{\ast}) \phi \right] + 2 e^2 |\phi|^2 {\bf A} = i \left[ \phi^{\ast} {\bf D} \phi - ({\bf D} \phi)^{\ast} \phi  \right], \label{curden}
\end{eqnarray}
where ${\bf D} = \nabla - i {\bf A}$ is the covariant derivative and the latter 
equation gives the electric current density. 

First, we give the solution of the vortex type, known as the 
Abrikosov-Nielsen-Olesen (ANO) vortex \cite{Abrikosov,Nielsen}.
By using $\phi = |\phi| e^{i \theta}$, Eq. (\ref{curden}) gives the 
vector potential as 
\begin{equation}
{\bf A} = \frac{\bf j}{2 e^2 |\phi|^2} + \nabla \theta.
\end{equation}
Without any current, the magnetic flux is quantized to be an integer 
($n=0,1,2,\cdots$) as, 
\begin{equation}
\int d^2 x B_z = \oint d {\bm \ell} \cdot  {\bf A}= \oint d {\bm \ell} \cdot \nabla \theta = 2 \pi n, 
\end{equation} 
where the vortex winding number is given by the first homotopy class: 
$n \in \pi_1 [{\rm U(1)}] = {\bf Z}$. 

\begin{figure}
\begin{center}
\includegraphics[width=0.5 \linewidth,keepaspectratio]{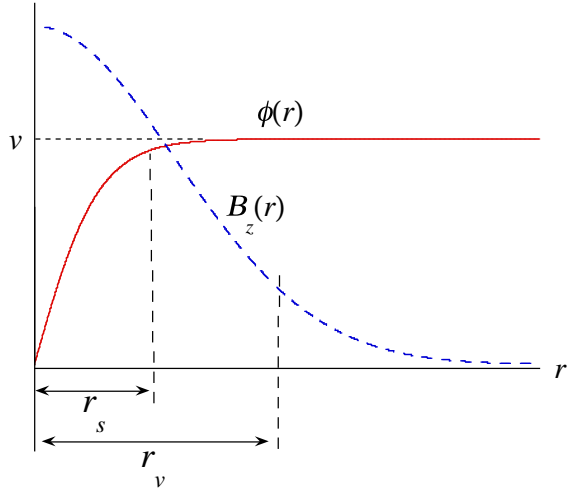}
\caption{Typical structure of the ANO vortex.} 
\label{ANOvortex}
\end{center}
\end{figure}
The structure of the ANO vortex solution is well known based on the numerical 
solution of Eq. (\ref{abehigseq}). The qualitative feature of the fields $\phi$ 
and $B_z$ is represented in Fig \ref{ANOvortex}. 
The Higgs field is reduced from the VEV at the vortex core, in which
U(1) gauge symmetry is recovered. The magnetic flux is concentrated 
around the vortex core. There are two length scales over which the above 
two fields vary spatially. The gauge mass $m_v \simeq \sqrt{2} e v$ and 
the scalar mass $m_s \simeq \sqrt{\lambda} v$ 
give the penetration depth $r_v = m_v^{-1} \simeq (\sqrt{2} e v)^{-1}$ 
and the coherence length $r_s = m_s^{-1} \simeq (\sqrt{\lambda} v)^{-1}$, respectively. 
The ratio $\beta = m_s^2/m_v^2 = \lambda/2e^2$ is the only free 
parameter in the model. 
For $r_v < r_s$ $(\beta < 1)$, the superconductor is classified as ``type I". 
Then, the interaction between vortices is attractive due to the exchange of the 
scalar fields and is consequently unstable under a magnetic field. 
For $r_v > r_s$ $(\beta > 1)$, the superconductor is classified as ``type II". The exchange of 
the gauge fields leads to a repulsive interaction between vortices and is stable under a magnetic field. 
The case $r_v = r_s$ $(\beta = 1)$ is known as the ``critical coupling", where 
there are no interactions between static vortices.  

For critical coupling ($\lambda=2e^2$), the equations for 
the topological defects can be simplified by taking the Bogomol'nyi-Prasad-Sommerfield (BPS) bound 
\cite{Bogomolnyi,Prasad} of the energy as 
\begin{eqnarray}
H &=& \int d^2 x \left[ \left| D_x \phi + i D_y \phi \right|^2 + \frac{1}{2e^2} \left\{B_z - e^2 (|\phi|^2 - v^2)^2 \right\}^2 \right] - v^2 \int d^2 x B_z \nonumber \\
&\geq& - v^2 \int d^2 x B_z = 2 \pi n v^2, \label{Bogobound}
\end{eqnarray}
where we have integrated in the uniform $z$-direction. 
When the equality holds, we can obtain the energy minimum and thus the most stable 
configuration for a fixed vortex number $n$. 
The minimized condition yields a set of simple linear equation known as the BPS equation 
\begin{equation} 
(D_x + i D_y) \phi = 0, \hspace{5mm} B_z = e^2 (|\phi|^2 - v^2) . 
\end{equation}
Fields satisfying the BPS equations are automatically minima of the energy 
for a given $n$, and hence are guaranteed to be stable. 

\subsection{Vortex in Abelian--two-Higgs model}\label{AbetwoHigtheory}
Next, we consider the Abelian--two-Higgs model \cite{Vachaspati}. 
The hamiltonian is given by the direct generalization of the Abelian-Higgs model where the 
complex scalar field is replaced by an SU(2) doublet $\Phi^{T}=(\Phi_1,\Phi_2)$ as 
\begin{eqnarray}
H = \int d^3 x \biggl[ \frac{1}{2e^{2}} ({\bf E}^{2} + {\bf B}^{2}) + |(\nabla -i {\bf A}) \Phi |^{2} 
 + \frac{\lambda}{4} (\Phi^{\dagger} \Phi - v^2 )^{2}  \biggr].  \label{AbeliantwoHiggs}
\end{eqnarray} 
This model is similar to the gauged two-component Ginzburg-Laudau model used to describe 
unconventional superconductors. 
This hamiltonian is invariant under the local U(1) gauge transformation ${\rm U(1)}_{\rm L} : \Phi \to e^{i \alpha({\bf r})} \Phi$ 
and ${\bf A} \to {\bf A} + \nabla \alpha({\bf r})$, 
and the global SU(2) operation ${\rm SU(2)}_{\rm G} : \Phi \to e^{i {\bm \gamma} \cdot {\bm \sigma} } \Phi$, 
where ${\bf \gamma}= \gamma {\bf n}$ with the positive constant 
$\gamma \in [ 0, 4 \pi ) $ and unit vector ${\bf n}$, and ${\bm \sigma}$ is the Pauli matrix.
Because the isotropy group is ${\rm U(1)}_{\rm L+G}$ \cite{tyuu3}, the order parameter space becomes 
$[{\rm U(1)}_{\rm L} \times {\rm SU(2)}_{\rm G}]/{\rm U(1)}_{\rm L+G} \simeq S^3$. 
The first homotopy group of the order parameter space is trivial $\pi_1(S^3) = 0$ (simply connected), 
so there are no topological vortex solutions. On the other hand, 
looking only at the gauged part of the symmetry, 
the local ${\rm U(1)}_{\rm L}$ symmetry actually breaks as in the case 
of the Abelian-Higgs model and thus $\pi_1[{\rm U(1)}_{\rm L}] ={\bf Z}$; 
the ${\rm U(1)}_{\rm L}$ gauge orbit away from the degenerate ${\rm SU(2)}_{\rm G}$ orbit 
is still expensive in terms of gradient energy, so it is possible that a vortex is stable. 
This vortex is called a {\it semilocal vortex}, defined in a non-topological manner \cite{Vachaspati}.

After the symmetry breaking, there are two Nambu Goldstone bosons, one scalar of mass 
$m_s \simeq \sqrt{\lambda} v$ and a massive vector boson of mass $m_v \simeq \sqrt{2} e v$, 
as in the Abelian-Higgs case. 
In the special case for critical coupling $\beta =\lambda/2e^2= 1$, one can take the BPS bound 
in the same way as for Eq. (\ref{Bogobound}): 
\begin{equation}
(D_x + i D_y) \Phi = 0, \hspace{5mm} B_z = e^2 (\Phi^{\dag}\Phi - v^2) . \label{BPSsemi}
\end{equation}
where the field $\phi$ is just replaced with $\Phi$. 
Thus, for the fixed topological sector, one can obtain the configuration satisfying the BPS 
equations, which is a local minimum of the energy and automatically stable, even though 
the vacuum manifold is simply connected. 
In this case, the vortex energy is independent of its core size, and thus the vortex can 
have an arbitrary size. 
For $\beta \neq 1$ the stability of the vortices depends on the dynamics and is controlled by the 
parameter $\beta$ \cite{James}. When the vortex energy is written in terms of its size, the energy monotonically 
increases as a function of the size for $\beta<1$ (type I case). Then, the vortex tends 
to shrink to zero size, which means that only one component of $\Phi$ survives while the 
other disappears, resulting in a stable ANO vortex. On the other hand, the energy becomes 
a monotonically decreasing function with respect to the size for $\beta > 1$ (type II case). 
In that case, the vortex is unstable for expansion and its winding eventually disappears. 

\begin{figure}
\begin{center}
\includegraphics[width=0.5 \linewidth,keepaspectratio]{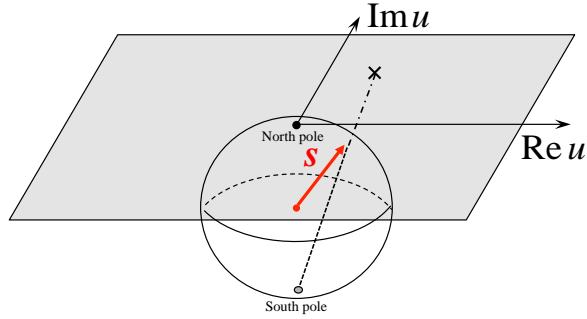}
\caption{Stereographic projection from the sphere to the tangent plane at the north pole.} 
\label{stereographic}
\end{center}
\end{figure}

\begin{figure}
\begin{center}
\includegraphics[width=0.4 \linewidth,keepaspectratio]{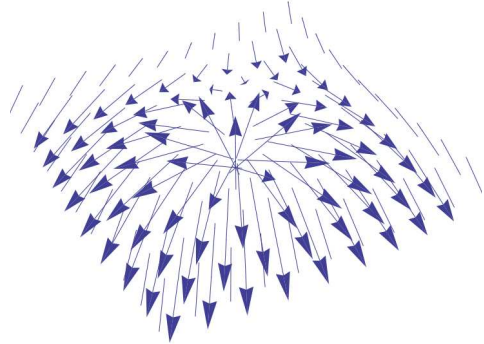}
\caption{Spin profile of a lump (2D skyrmion, coreless vortex). The spin orients upwards
at the center and it continuously rotates from up to down as it moves outward radially.} 
\label{lump}
\end{center}
\end{figure}

For $e \to \infty$, the electromagnetic 
energy (the first term of Eq. (\ref{AbeliantwoHiggs})) vanishes, so that the vector potential ${\bf A}$ becomes 
non-dynamical variables, given by 
\begin{equation}
{\bf A} = -\frac{i}{2}  \frac{\Phi^{\dag} \nabla \Phi - (\nabla \Phi^{\dag}) \Phi}{\Phi^{\dag} \Phi}. 
\label{vectorpotphi}
\end{equation}
In addition, for the limit $\lambda \to \infty$, the radial degree of freedom of $\Phi$ is frozen 
and the minimum of the potential yields the VEV 
$\Phi^{\dagger} \Phi = |\Phi_1|^2 +|\Phi_2|^2 = v^2$. 
As seen below, the order parameter space can be 
reduced to $S^3/{\rm U(1)}_{\rm L} \simeq S^2$ (sphere), and the corresponding model 
is also reduced to the O(3) NL$\sigma$M ({\bf C}P$^1$ model). 
The order parameter can be written with the stereographic coordinate $u$ by 
\begin{equation}
\Phi = \frac{v}{\sqrt{1+ |u|^{2}}} 
\left(
\begin{array}{c}
1 \\
u 
\end{array}
\right)
\label{Phiutrans}
\end{equation}
and the corresponding unit isovector (pseudospin) ${\bf s}$ can be given as
\begin{eqnarray}
{\bf s} = \frac{\Phi^{\dagger} {\bm \sigma} \Phi}{\Phi^{\dagger} \Phi} = \left( \frac{u+u^\ast}{1+|u|^2} ,  
-i \frac{u-u^\ast}{1+|u|^2},  \frac{1- |u|^{2}}{1+|u|^2}  \right), \label{spindefeinition}
\end{eqnarray}
where $u=0$ and $\infty$ corresponds to $\Phi^T=(v,0)$ and $\Phi^T=(0,v)$, respectively, and, in terms of 
the pseudospin, it corresponds to the north and south pole, respectively, as shown in Fig. \ref{stereographic}. 
The hamiltonian Eq. (\ref{AbeliantwoHiggs}) is reduced to 
\begin{eqnarray}
H =\int d^3 x \frac{(\nabla u^{\ast}) (\nabla u)}{(1+|u|^2)^2} = \frac{1}{4} \int d^3 x \sum_{\alpha} (\nabla s_{\alpha})^2 
\end{eqnarray}
In this limit, the semilocal vortex is reduced to a coreless vortex, known 
as a lump (2D skyrmion) \cite{Belavin}; its spin configuration is shown in Fig. \ref{lump}. 
The second homotopy group of $S^2$ is nontrivial 
as $\pi_2(S^2) = \pi_2[{\rm SU(2)}/{\rm U(1)}] \simeq \pi_1[{\rm U(1)}] \simeq {\bf Z}$, which gives a 
2D skyrmion charge; the hedgehog configuration of the pseudospin can 
be mapped to the configuration of the 2D skyrmion through the stereographic projection. 

\subsection{D-brane solitons in the massive Abelian--two-Higgs model}\label{AbetwomassHigtheory}
In the following, we assume a critical coupling $\lambda = 2 e^{2}$. 
This regime allows us to make an analytical treatment with the BPS equation as shown above.  
Our main target is to analyze D-brane solitons, which consist of vortices and 
domain walls. However, since the order parameter space of the model Eq. (\ref{AbeliantwoHiggs}) 
is $S^3$, there cannot be a domain wall. To realize a discrete ground state configuration 
in space, the mass for the Higgs fields should be introduced. 
We extend the model Eq. (\ref{AbeliantwoHiggs}) 
to the massive Abelian--two-Higgs model, given by 
\begin{eqnarray}
H = \int d^3 x \biggl[ \frac{1}{2e^{2}} ({\bf E}^{2} + {\bf B}^{2}) + |(\nabla -i {\bf A}) \Phi |^{2} 
+ \frac{1}{2e^{2}} ( \nabla \Sigma)^{2} \nonumber \\
+ \frac{e^{2}}{2} (\Phi^{\dagger} \Phi - v^2 )^{2} 
 + \Phi^{\dagger} ( \Sigma {\bf I} - {\bf M})^{2} \Phi  \biggr].
 \label{multihiggsmassive}
\end{eqnarray}
Here, we have introduced the neutral scalar field $\Sigma$, 
$2 \times 2$ identical matrix ${\bf I}$ 
and the mass matrix 
\begin{eqnarray}
{\bf M} = \left(
\begin{array}{ll}
M_1 & 0 \\
0 & M_2
\end{array}
\right).
\end{eqnarray}
The role of the neutral scalar field $\Sigma$ is to generate different vacua or, in other words, to determine 
the direction of the symmetry breaking from ${\rm SU(2)}_{G}$ to ${\rm U(1)}_{G}$ of the $\Phi$ field; 
for example, see \cite{Shifman1,Eto:2006pg}. 
Actually, there are two ground states 
as $\langle \Phi \rangle^{T} = (v,0)$, $\langle \Sigma \rangle = M_1$, and 
$\langle \Phi \rangle^{T}  = (0,v)$, $\langle \Sigma \rangle = M_2$. 
Thus, one can make a domain wall between these two vacua. 
For a single domain wall, the BPS bound is given as 
\begin{eqnarray}
H = \int dz \biggl\{ | D_{z} \Phi + (\Sigma {\bf I} - {\bf M}) \Phi |^{2} 
+ \frac{1}{2e^{2}} \left[ B_{z} - \partial_{z} \Sigma - e^{2} (|\Phi|^{2} - v^2 )^{2} \right]^{2} \nonumber \\ 
+ t_w + \partial_z J_z  \biggr\}
\label{BPSwallbound}
\end{eqnarray}
with the topological charge density 
\begin{equation}
t_w = v^2 \partial_z \Sigma
\end{equation}
and the current 
\begin{equation}
J_z = - \Phi^{\dag} (\Sigma {\bf I} - {\bf M}) \Phi
\end{equation}
After the integration, the contribution of the current vanishes and 
the total topological charge (tension) of the domain wall is 
\begin{equation}
T_w = \int dz t_w = v^2 \Delta M, 
\end{equation}
associated with the difference of the mass $\Delta M = M_1 - M_2$ between the two vacua. 

For $e \to \infty$, both the vector potential ${\bf A}$ and the scalar field $\Sigma$ are 
no longer independent dynamical fields and their form is determined 
by the minimization of the energy with respect to ${\bf A}$ and $\Sigma$, where ${\bf A}$ is given 
by Eq. (\ref{vectorpotphi}) and $\Sigma$ is 
\begin{equation}
\Sigma = \frac{\Phi^{\dag} {\bf M} \Phi}{\Phi^{\dag} \Phi} = \frac{M_1 |\Phi_1|^2 + M_2 |\Phi_2|^2}{|\Phi_1|^2+|\Phi_2|^2}.
\end{equation}
When we choose $M_1=M/2$ and $M_2=-M/2$, $\Sigma$ is identical to the $z$-component 
of the pseudospin field: 
\begin{equation}
\Sigma = \frac{M}{2} \frac{|\Phi_1|^2 - |\Phi_2|^2}{|\Phi_1|^2+|\Phi_2|^2} = \frac{M}{2} s_{z}. 
\label{SigmaMrelation}
\end{equation}
Since the limit $e \to \infty$ means $\lambda \to \infty$ here, the model also reduces 
to the massive O(3) NL$\sigma$M with the relations Eqs. (\ref{Phiutrans}) and (\ref{spindefeinition}). 
The hamiltonian can be rewritten as 
\begin{eqnarray}
H &=& \frac{1}{4} \int d^3 x \left[ \sum_{\alpha} (\nabla s_{\alpha})^2 + M^2 (1-s_z^2) \right] \nonumber \\
&= & \int d^3 x \frac{(\nabla u^{\ast}) (\nabla u) + M^2 |u|^2}{(1+|u|^2)^2}. \label{massiveNLsm}
\end{eqnarray}
The exact solutions of the BPS equation can be found to be \cite{Gauntlett}
\begin{equation}
\Phi = \frac{v}{\sqrt{1+e^{-2 M (z-z_0)}}} 
\left(
\begin{array}{c}
1 \\
e^{- M (z-z_0) - i \theta_0}
\end{array}
\right)
\end{equation}
Here, $z_0$ represents the position of the flat domain wall ($s_z = 0$) whose 
transverse shift causes the Nambu-Goldstone mode due to 
breaking of the translational invariance. The phase $\theta_0$ corresponds 
to the azimuthal angle of the pseudospin ${\bf s}$, 
causing the breaking of the global U(1) locally along the wall. 
By promoting these two variables to dynamical 
fields, we can construct an effective theory of the domain wall. 
The low-energy dynamics of a single domain wall is simply described by the DBI action \cite{Gauntlett,Shifman1}, 
where the {\it local} U(1) gauge fields living on the wall are created by the dual transformation of the 
localized zero mode of the phase $\theta_0$ due to the breaking of the {\it global} U(1) in the original system. 
This is why this domain wall can be 
identified as an analog of a D-brane, as stated in Sec.\ref{Intro}
\begin{figure}
\begin{center}
\includegraphics[width=0.85 \linewidth,keepaspectratio]{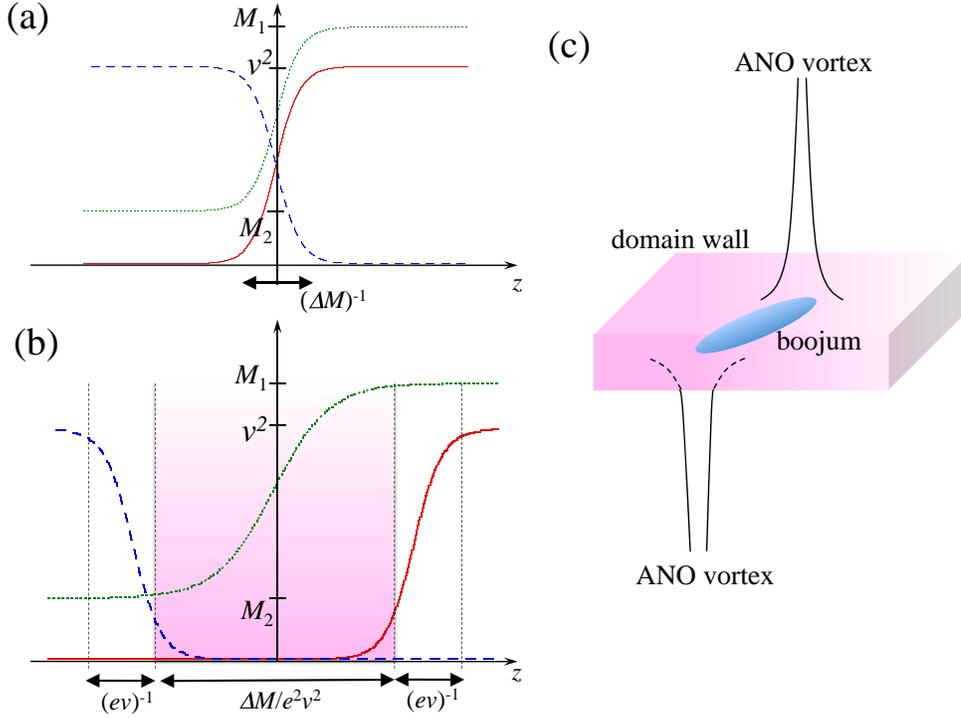}
\caption{Typical examples of a domain wall and a wall-vortex complex (D-brane soliton). 
The profile of the domain wall for (a) the $e \to \infty$ limit and (b) finite $e$, where 
$\Phi_1$, $\Phi_2$, and $\Sigma$ are represented by solid, dashed, and dotted curves, respectively. 
Figure (c) represents the typical configuration of a D-brane soliton for finite $e$. Two vortices 
end on the domain wall from both sides.} 
\label{domainwall}
\end{center}
\end{figure}

The structure of the domain wall profile at finite $e$ depends on the 
dimensionless parameter $e^2 v^2/\Delta M^2$ \cite{Shifman1}. A typical profile of the domain wall is 
shown in Fig. \ref{domainwall}. For the sigma model (strong gauge coupling) limit 
$e^2 v^2 \gg \Delta M^2$ all fields interpolate between their vacuum values over 
a single length scale $(\Delta M)^{-1}$ [see Fig. \ref{domainwall}(a)]. 
In the opposite weak gauge coupling limit $e^2 v^2 \ll \Delta M^2$, 
the domain wall has a three-layer structure, as shown Fig. \ref{domainwall}(b). 
In the two outer layers, each field $\Phi_i$ drops exponentially to zero with the width $(ev)^{-1}$. In the 
inner layer which has width $\Delta M/e^2 v^2$, the adjoint field $\Sigma$ interpolates 
between its two vacuum values. 

Next, we consider the composite soliton that consists of the vortices and the domain walls.
Combining Eqs. (\ref{Bogobound}) and (\ref{BPSwallbound}), 
the BPS bound under the fixed topological sector is given by 
\begin{eqnarray}
H = \int d^3 x \biggl \{ | D_{z} \Phi + (\Sigma {\bf I} - {\bf M}) \Phi |^{2} 
+ |D_{x} \Phi + i D_{y} \Phi|^{2} \nonumber \\
+ \frac{1}{2e^{2}} \biggl[ |B_{x} - \partial_{x} \Sigma|^{2} + |B_{y} - \partial_{y} \Sigma|^{2} 
+ \left\{ B_{z} - \partial_{z} \Sigma - e^{2} (|\Phi|^{2} - v^2 )^{2} \right\}^{2} \biggr] \nonumber \\
+ t_{w} + t_{v} + t_{m} + \nabla \cdot {\bf J} \biggr\}
\end{eqnarray}
with the topological charge densities 
\begin{eqnarray}
t_{w} = v^2 \partial_{z} \Sigma, \\
t_{v} = - v^2 B_{z}, \\
t_{m} = \frac{1}{e^{2}} {\bf B} \cdot \nabla \Sigma =  \frac{1}{e^{2}} \nabla \cdot (\Sigma {\bf B}), \label{monochargeden}
\end{eqnarray}
and the current 
\begin{equation}
{\bf J} = \left( i \Phi^{\dagger} D_{y} \Phi, 
- i \Phi^{\dagger} D_{x} \Phi, - \Phi^{\dagger} (\Sigma {\bf I} - {\bf M}) \Phi \right). 
\end{equation}
An important result is the appearance of the ``monopole charge" $t_m$ 
of Eq. (\ref{monochargeden}). We can write ${\bf B} \cdot \nabla \Sigma $ as $ \nabla \cdot ({\bf B} \Sigma)$ 
where $\nabla \cdot {\bf B} = 0$ holds. Note that the charge is defined by the {\it projected} magnetic field $\Sigma {\bf B}$. 
This monopole should be distinguished from an usual hedgehog of $S^2$, because 
it cannot exist in U(1) gauge theory. This point-like defect is expected 
to be located on the domain wall as the connecting point of a domain wall and a vortex, 
so that this charge may be referred to as a ``boojum charge". 
The topological charge is given by integrating the topological charge 
density in the corresponding spatial dimension as 
\begin{eqnarray}
T_{w} = \int dz t_w = v^2 (M_1-M_2), \\
T_{v} = \int d^{2}x  t_v = 2 \pi n v^2, \\
T_{m} =  \int d^{3}x  t_m = \frac{1}{e^{2}} \int d^{3}x \nabla \cdot (\Sigma {\bf B}). \label{monocharge}
\end{eqnarray}
The typical configuration of a D-brane soliton is depicted in Fig. \ref{domainwall}(c). 
Since there is only one component of $\Phi$ on either side of the wall, the vortices should be 
identified as ANO vortices. The boojums are expected to be spread inside the domain wall 
on which the two vortices end from both sides, as shown in Fig. \ref{domainwall}(c).

\begin{figure}
\begin{center}
\includegraphics[width=0.85 \linewidth,keepaspectratio]{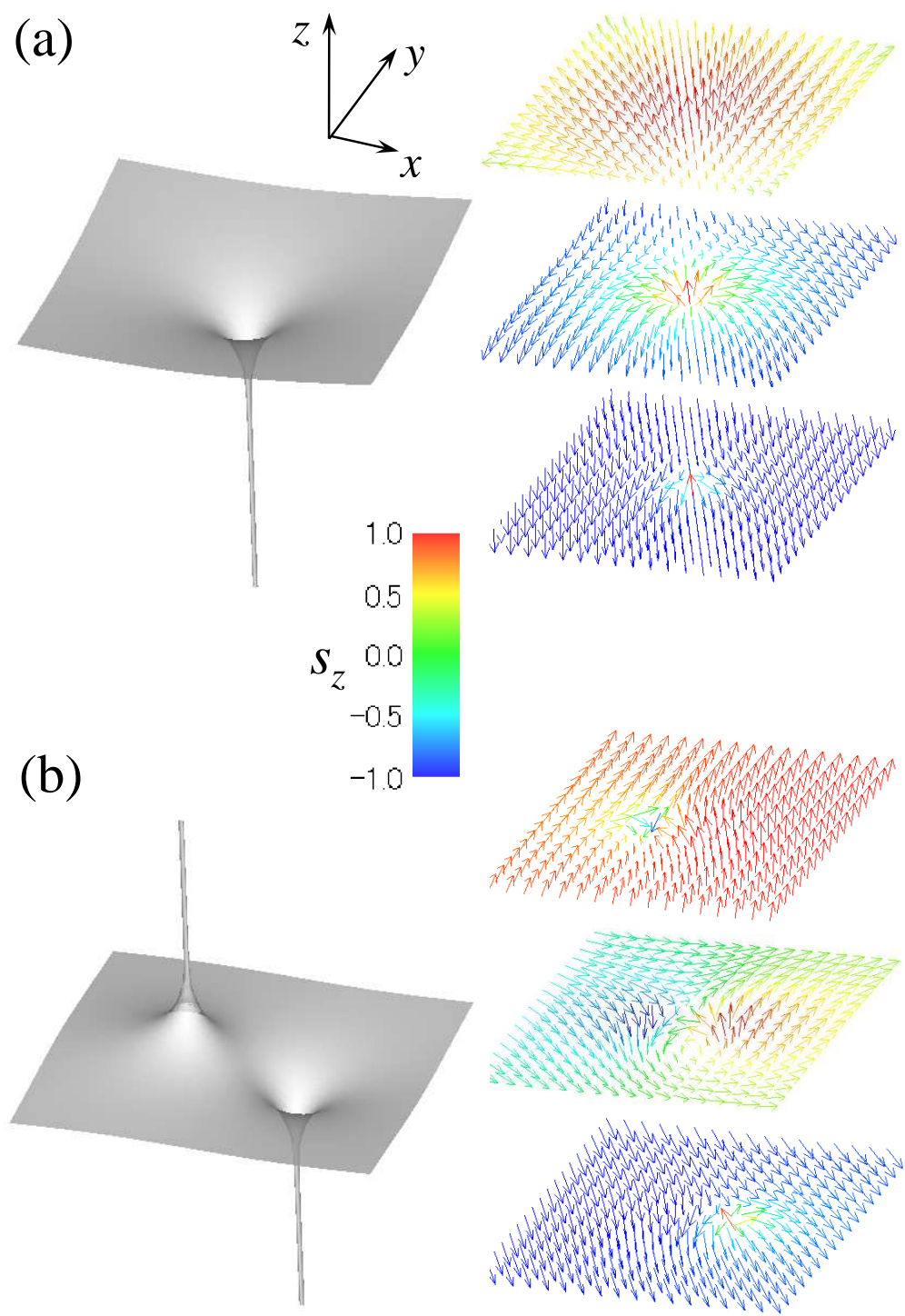}
\caption{Typical configuration of a D-brane soliton in the field theoretical model under the sigma model limit. 
The left pictures show the isosurface of the domain wall $u=1$ ($s_z=0$). 
(a) A single vortex is attached to the domain wall from one side. (b) Two vortices 
are attached to the wall from both sides. The spin configuration in three $z=$const planes, 
chosen as $z>0$, $z=0$, and $z<0$, are also depicted on the right.} 
\label{dbranesolifig}
\end{center}
\end{figure}

In the strong gauge coupling limit, the analytic solution for the BPS wall-vortex soliton can also be derived as \cite{Isozumi} 
\begin{equation}
\Phi (w, z) = \frac{v}{\sqrt{H_{0}(w)^{\dag} e^{2{\bf M} z} H_{0} (w)}} H_{0}(w) e^{{\bf M} z}. \label{exactsolution}
\end{equation}
Here, $H_0(w)$ is the Moduli matrix given by 
\begin{equation}
H_{0}(w) = 
\left(
\begin{array}{c}
a_{1} \prod_{j=1}^{k_{1}} (w-w_{j}^{(1)}) \\
a_{2} \prod_{j=1}^{k_{2}} (w-w_{j}^{(2)})
\end{array}
\right).
\end{equation}
In this notation, the $k_i$ individual vortices are positioned at $w=w_{j}^{(i)}$ in the $\Phi_i$ domain ($i=1,2$). 
For ${\bf M}={\rm diag.} (M/2, -M/2)$, the solution can be written as
\begin{eqnarray}
\Phi(w,z) = \frac{v}{\sqrt{1 + e^{-2 M (z-z_0)} |Z|^2}} 
\left(
\begin{array}{c}
1 \\
e^{-M(z-z_0)+i \theta_0} Z
\end{array}
\right).
\end{eqnarray}
Here, the holomorphic function $Z=Z(w)$ represents the contribution of 
the coreless vortex (lump): 
\begin{equation}
Z(w) =\frac{\prod_{j=1}^{k_2} (w-w_{j}^{(2)})}{\prod_{j=1}^{k_1} (w-w_{j}^{(1)})}.
\end{equation} 
This form of the solution is equivalent to that found by Gauntlett {\it et al.} \cite{Gauntlett}, 
where $u(w,z) = e^{-M (z-z_0) + i \theta_0}Z(w)$ by using the transformation Eq. (\ref{Phiutrans}), and 
they consider the case $k_{2}=k_{1}-1$; Isozumi {\it et al.}, on the other hand, consider 
the more general case including $k_1=k_2$ \cite{Isozumi}. 
We can thus construct solutions in which an arbitrary number of vortices are connected to
the domain wall. In the BPS solution, the energy is independent of the vortex
positions $w_j^{(1,2)}$ on the domain wall; in other words, there
is no static interaction between vortices.
The structure of a D-brane soliton with a simple wall-vortex configuration is shown in Fig. \ref{dbranesolifig}. 
Here, the $\Phi_1$ and $\Phi_2$ fields are positioned in the $z>0$ and $z<0$ regions, respectively. 
In Fig. \ref{dbranesolifig}(a), a single vortex (lump) exists in $z < 0$ ($\Phi_2$ domain), given by $Z=w$. 

The edge of the vortex attaches to the wall, causing it to bend 
logarithmically as $z = \log |w| / M$. Figure \ref{dbranesolifig}(b) shows a solution in which
both fields have one vortex connected to the wall, corresponding to $Z=(w-x_0^{(2)})/(w-x_0^{(1)})$. 
In this case, the wall becomes asymptotically flat due to the balance of the vortex tension. 

As shown above, the connecting points of the wall and vortices can give rise to point defects known 
as boojums \cite{Isozumi,Sakai,Auzzi}. 
The presence of a boojum is confirmed by calculating the boojum charge by Eq. (\ref{monocharge}). 
Because the domain wall is logarithmically bent when a single vortex is attached to one side, 
this situation does not give a clear integral region to calculate the topological charge. 
Let us consider that two vortices with winding number $n=1$ are attached to both sides of the wall. 
Then, we do not need to take care of the integral region of Eq. (\ref{monocharge}) because of the asymptotically flat wall. 
The charge $T_m$ in this configuration, i.e., the charge of two boojums, can be calculated as 
\begin{eqnarray}
T_m &=& \frac{1}{e^{2}} \int d^{3}x \nabla \cdot (\Sigma {\bf B})  \nonumber \\ 
&=& \frac{1}{e^{2}} \left( \left[ \Sigma \int d^2 x B_z \right]_{z=\infty} - \left[ \Sigma \int d^2 x B_z \right]_{z=-\infty} \right) \nonumber \\
&=& \frac{1}{e^{2}} \left( \left[ \Sigma T_v \right]_{z=\infty} - \left[ \Sigma T_v \right]_{z=-\infty} \right) \nonumber \\
&=& - \frac{2 \pi}{e^{2}} (M_1 - M_2). 
\label{boochacal}
\end{eqnarray}
It turns out that boojums have negative monopole charge and negative energy, 
despite the signs of the charges of the vortices and domain walls \cite{Sakai}. 
Since one monopole is characterized by $T_{m} = \frac{2 \pi}{e^2} (M_1 - M_2) > 0$ \cite{Hooft}, 
one boojum has a half of the single negative monopole charge. 
It has been proposed that the boojum plays a role in binding the wall 
and the vortex, because the energy of this composite soliton is smaller 
than the independent sum of the energies of the wall and the vortices. 
 
\section{D-brane solitons in two-component BECs}\label{formulation}
\begin{figure}
\begin{center}
\includegraphics[width=0.85 \linewidth,keepaspectratio]{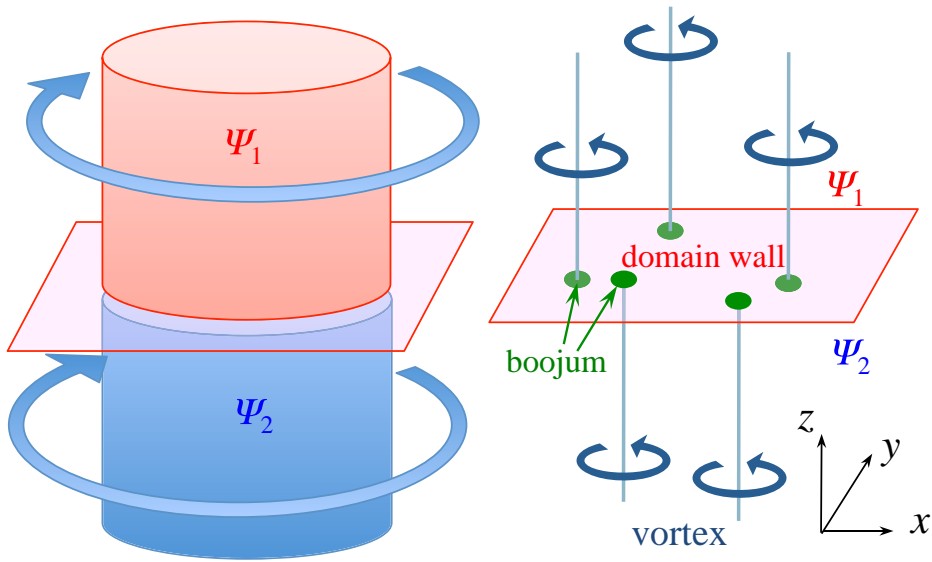}
\caption{Schematic illustration of the wall-vortex soliton 
configuration in two-component BECs. 
The two-component BECs $\Psi_{1}$ ($z>0$) and $\Psi_{2}$ ($z<0$) are 
separated by a domain wall near the $z=0$ plane. 
The domain wall can be defined as a plane on which both components have the same amplitude. 
We assume that vortex lines, nucleated by rotation around the $z$-axis, are 
perpendicular to the wall. } 
\label{dbraneBECponch}
\end{center}
\end{figure}

Based on the above analysis, we consider a system of cold atomic two-component BECs and study
the detailed properties of the wall-vortex complex (D-brane soliton) in this system. 
The system under consideration is schematically shown in Fig. \ref{dbraneBECponch}.
Two-component BECs have been realized by using a mixture of atoms with two 
hyperfine states of $^{87}$Rb \cite{Hall,Mertes,Tojo} 
or a mixture of two different species of atoms such 
as $^{87}$Rb-$^{41}$K \cite{Modugno,Thalhammer}, 
$^{85}$Rb-$^{87}$Rb \cite{Papp}, and $^{87}$Rb-$^{133}$Cs \cite{McCarron}. 
The experiments in Refs.\cite{Tojo,Papp} demonstrated that the miscibility 
and immiscibility can be controlled by tuning the atom-atom 
interaction via Feshbach resonances. 
Here, the domain wall is referred to as a boundary of 
phase-separated two-component BECs and is well-defined as a plane 
on which both components have the same amplitudes \cite{Tim,Ao,Trippenbach,Barankov,Schaeybroeck}. 
The vortices can be arranged by applying a rotation to the BEC 
around the $z$-axis or imprinting an appropriate phase profile to 
the condensate with engineered atom-laser coupling \cite{Fetterreview}. 
 
\subsection{Gross-Pitaevskii model}
Two-component BECs are described by the order parameters 
\begin{equation}
\Psi = 
\left(
\begin{array}{c}
\Psi_1 \\
\Psi_2
\end{array}
\right)
=
\left(
\begin{array}{c}
\sqrt{\rho_1} e^{i \theta_1} \\
\sqrt{\rho_2} e^{i \theta_2}
\end{array}
\right), 
\end{equation}
which are condensate wave functions with density $\rho_j$ 
and phase $\theta_j$ ($j=1,2$). 
They are typically confined in a trapping potential $V_j$. 
The energy functional of the two-component Gross-Pitaevskii (GP) model is given by
\begin{eqnarray}
H [\Psi_{1},\Psi_{2}] = \int d^3 x \biggl\{ \sum_{j = 1,2} 
\biggl[ \frac{\hbar^{2}}{2m_{j}}  \left| \left(\nabla - i \frac{2m_j}{\hbar} \tilde{\bf A} \right) \Psi_{j} \right|^{2} 
+ (V_j - \mu_{j}) |\Psi_{j}|^{2} \nonumber \\ 
+ \frac{g_{jj}}{2} |\Psi_{j}|^{4} \biggr]  + g_{12} |\Psi_{1}|^{2} |\Psi_{2}|^{2} \biggr\}.   
\label{energyfunctio2}
\end{eqnarray}
Here, $m_{j}$ is the particle mass of the $j$-th component and $\mu_j$ is its chemical potential. 
The coefficients $g_{11}$, $g_{22}$, and $g_{12}$ represent the 
atom-atom interactions and are expressed as
\begin{equation}
g_{jk} = \frac{2 \pi \hbar^{2} a_{jk}}{m_{jk}}, \hspace{5mm} \frac{1}{m_{jk}} = \frac{1}{m_{j}} + \frac{1}{m_{k}}
\end{equation} 
in terms of the s-wave scattering lengths $a_{11}$ and $a_{22}$ 
between atoms in the same component and $a_{12}$ between atoms 
in different components. The s-wave scattering length can be 
tuned by the Feshbach resonance \cite{Tojo,Papp}. 
The most striking difference between 
the GP model and the Abelian-Higgs model is that the gauge field is not 
a dynamical variable because the atoms in the condensate are electrically neutral. 
The vector potential $\tilde{\bf A}$ is just an external field and is generated by 
(i) the rotation of the system $\tilde{\bf A} = ({\bf \Omega} \times {\bf r}) / 2 $ \cite{Fetterreview} 
or (ii) a synthesis of the artificial magnetic field by the laser-induced Raman coupling between 
the internal hyperfine states of the atoms \cite{Lin}. 

To show the similarity of the formulation between the two-component BECs and the 
field theories in the previous section, we consider the simple case of 
a homogeneous system without a trapping potential $V_j=0$. 
By setting $m_1 = m_2 = m$ and $g_{11} = g_{22} = g$, the potential in 
the hamiltonian Eq. (\ref{energyfunctio2}) can be written as 
\begin{eqnarray}
V(\Psi) = \frac{g}{2} \left( \Psi^{\dag} \Psi - \frac{\bar{\mu}}{g} \right)^2 
+ (g_{12} - g) |\Psi_1|^2 |\Psi_2|^2 \nonumber \\
- \frac{\mu_1 - \mu_2}{2} (|\Psi_1|^2 - |\Psi_2|^2), \label{potentialPsi}
\end{eqnarray}
where $\bar{\mu} = (\mu_1 + \mu_2)/2$, and 
the constant term has been removed. 
For $g_{12}=g$ and $\mu_1 = \mu_2$, the hamiltonian has a global U(2) symmetry. 
When these conditions are not satisfied, the U(2) symmetry is broken to U(1)$\times$U(1). 
Note that for $g_{12} < g$, the second term in Eq. (\ref{potentialPsi}) implies that the 
two components prefer to overlap spatially, while for $g_{12} > g$ the condensates tend to 
be segregated to reduce the overlap region. 
By introducing the characteristic length scale
\begin{equation}
\xi = \frac{\hbar}{\sqrt{2 m \bar{\mu}}},
\end{equation}
the hamiltonian can be written in a dimensionless form
\begin{eqnarray}
H = \int d^3 x \biggl[ |(\nabla - i \tilde{\bf A}) \Psi|^2 + \frac{\lambda}{4} \left( \Psi^{\dag} \Psi - v^2 \right)^2 
+ \Delta g |\Psi_1|^2 |\Psi_2|^2 \nonumber \\ 
- \delta \mu (|\Psi_1|^2 - |\Psi_2|^2) \biggr]
\label{dimlessenergy}
\end{eqnarray}
with the dimensionless parameters $\lambda = 2g/\bar{\mu} \xi^3$, $v^2 = \bar{\mu} \xi^3 / g$, 
$\Delta g = (g_{12} - g)/\xi^3 \bar{\mu}$, and $\delta \mu = (\mu_1-\mu_2)/\bar{\mu}$. 
For $\delta \mu = 0$, Eq.(\ref{dimlessenergy}) is a similar model, discussed by Shifman and Yung 
as a ``toy model" \cite{Shifman1}. The difference from this toy model is the absence of the electromagnetic energy, 
which suggests that the GP model is a counterpart of the Abelian-Higgs model in the 
limit $e \to \infty$ but for finite $\lambda$. 

\subsection{Vortices in two-component BECs} 
We first briefly describe the vortex states in two-component BECs; see Ref.\cite{KTUreview} for the details. 
We confine ourselves to the case $\Delta g = 0$, i.e. the SU(2) symmetric case for $\delta \mu = 0$ 
and the axisymmetric structure of the vortex states. Since this structure has been studied very well, 
we only consider the analogy with the vortices found in the Abelian--two-Higgs model. 

\begin{figure}
\begin{center}
\includegraphics[width=0.98 \linewidth,keepaspectratio]{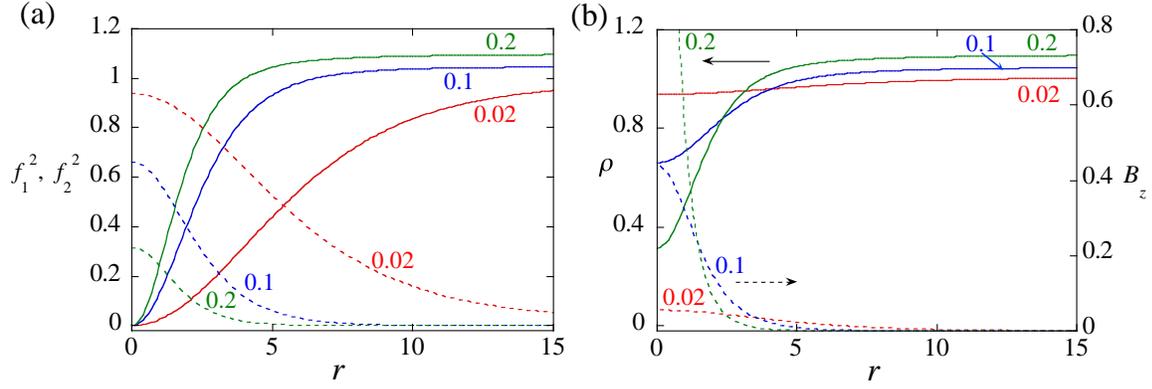}
\caption{Axisymmetric vortex structure of two-component BECs for vortex 
winding number $(q_{1}, q_{2}) = (1,0)$. The parameter values are $\lambda = 2$ 
and $\Delta g = 0$ in Eq. (\ref{dimlessenergy}). The U(2)-symmetry breaking 
parameter $\delta \mu$ is chosen to be 0.02, 0.1, and 0.2.  
(a) Radial profile of $|\Psi_1|^2$ (solid curve) and $|\Psi_2|^2$ (dashed curve). 
(b) Total density $\rho$  (solid curve) and effective magnetic field $B_z$ (dashed curve).} 
\label{vortexfig}
\end{center}
\end{figure}

The solution of the vortex state can be obtained from 
the analysis of the coupled GP equations derived from Eq. (\ref{dimlessenergy}). 
The axisymmetric (singly-charged) vortex states are characterized by 
$\Psi_{i}({\bf r})= f_{i}(r) e^{i q_{i} \theta}$ and we consider the most simple case 
$(q_{1}, q_{2}) = (1,0)$. 
By applying the boundary condition $f_1(r=0) = 0$, $f_1(r=\infty) = v$, $f_2'(r=0) = 0$, 
and $f_2(r=\infty) = 0$, we can obtain a vortex structure consisting of the circulating $\Psi_1$
component that surrounds the non-rotating $\Psi_2$ component at the center, as shown in Fig. \ref{vortexfig}. 
Since the total density does not vanish at the vortex 
core, this vortex can be called a {\it coreless} vortex. 
If the parameter $\delta \mu$ is zero, the system does not allow vortices to exist 
since the vacuum manifold is $S^3$ which does 
not have non-contractible loops. Thus, the size of the vortex can be arbitrary; 
the core of the vortex actually fills the entire space, in
which case the meaning of the vortex is completely lost. 
Since the gauge field is absent in the system, there is no intrinsic mechanism to 
stabilize the vortex in a topological manner like a semilocal vortex in the Abelian--two-Higgs model. 
Actually, an extrinsic mechanism such as the rotation $\tilde{\bf A}$ or the trapping 
potential $V_j$ allows a stable coreless vortex even for $\delta \mu=0$ \cite{KTUreview}. 

When $\delta \mu > 0$ $(<0)$, the SU(2) symmetry of the system is explicitly broken to 
U(1) $\times$ U(1) and the order parameter space is 
[U(1)$\times$ U(1)]$/$U(1) = U(1) $\simeq S^1$ with the ground state 
$|\Psi_1| = v$ and $|\Psi_2| = 0$ ($|\Psi_1| = 0$ and $|\Psi_2| = v$), leading to the formation 
of stable vortex solutions in the order parameter $\Psi_1$ ($\Psi_2$). 
However, for smaller values of $|\delta \mu|$, $\Psi_2$ ($\Psi_1$) can
survive only around the vortex core. 
With increasing $|\delta \mu|$, the vortex core 
shrinks by an accompanying decrease in the population of $\Psi_2$  ($\Psi_1$) at the core. 
For large values of $|\delta \mu|$, only the conventional singular-type vortex of $\Psi_1$ remains 
and the $\Psi_2$ field vanishes everywhere in space; the symmetry is
restored in the vortex core and the core size is fixed by the
standard healing length $\xi = \hbar/\sqrt{2 m \mu_{1(2)}}$. 

Although the real gauge field is absent, we can define the {\it effective} 
vector potential by noting that the gauge field is coupled with the 
configuration of $\Psi$ in the limit $e \to \infty$ as seen in Eq. (\ref{vectorpotphi}): 
\begin{equation}
{\bf A}_{\rm eff} =  -\frac{i}{2}  \frac{\Psi^{\dag} \nabla \Psi - (\nabla \Psi^{\dag}) \Psi}{\Psi^{\dag} \Psi} = {\bf v}_s.
\label{vectorvelo}
\end{equation}
In terms of the neutral superfluid, this effective vector potential is equivalent 
to the superfluid velocity field ${\bf v}_s$. 
In addition, we can define the effective magnetic field through the relation 
\begin{equation}
{\bf B}_{\rm eff} = \nabla \times {\bf A}_{\rm eff} = {\bm \omega}_s, 
\end{equation}
which corresponds to the 
vorticity ${\bm \omega}_s =  \nabla \times {\bf v}_s$. 
Figure \ref{vortexfig} also depicts the distribution of $B_{{\rm eff} z}$. 
In the coreless vortex, the vorticity is continuously distributed 
around the vortex core at which $|{\bf v}|$ vanishes. 
As $\delta \mu$ increases, the vorticity shrinks together with the $\Psi_2$-component. 
In the limit of $\Psi_2 \to 0$,  the vortex becomes a singular type and the vorticity is concentrated 
on the origin as a delta function type, 
whose behavior is similar to the $B_z$ of the ANO vortex in the $e \to \infty$ limit. 

The behavior of the vorticity distribution can be understood as follows. 
The velocity field of Eq. (\ref{vectorvelo}) can be written as  
\begin{equation}
{\bf v}_s = \nabla \Theta - \cos \theta \nabla \varphi, \label{velophaserep}
\end{equation}
which consists of two parts: the current due to the global U(1) phase $\Theta = \theta_1 + \theta_2$ 
and that due to the variation of the polar angle $\theta$ and 
the azimuthal angle $\varphi = \theta_2 - \theta_1$ of the unit sphere ($S^2$). 
Here, the angles of $S^2$ can be related to the pseudospin field 
\begin{eqnarray}
{\bf s}&=&\frac{\Psi^{\dagger}  \bm{\sigma}  \Psi}{\Psi^{\dagger} \Psi}
= \left( \frac{\Psi_{1}^{\ast} \Psi_{2} + \Psi_{2}^{\ast} \Psi_{1}}{|\Psi_{1}|^2+|\Psi_{2}|^2},
-i \frac{\Psi_{1}^{\ast} \Psi_{2} - \Psi_{2}^{\ast} \Psi_{1}}{|\Psi_{1}|^2+|\Psi_{2}|^2}, 
\frac{|\Psi_{1}|^{2} - |\Psi_{2}|^{2}}{|\Psi_{1}|^2+|\Psi_{2}|^2}
\right) \nonumber \\
&=& \left( \sin \theta \cos \varphi ,
\sin \theta \sin \varphi ,
\cos \theta  \right) . \label{spindef}
\end{eqnarray}
From Eq. (\ref{velophaserep}), the vorticity can be written as 
\begin{eqnarray}
{\bm \omega}_s
= \biggl[ \nabla \times (\nabla \Theta) - \cos \theta \nabla \times (\nabla \varphi) 
+ (\nabla \theta) \times ({\sin} \theta \nabla \varphi) \biggr].
\label{vorticityeff}
\end{eqnarray}
The first and second terms in Eq. (\ref{vorticityeff}) vanish except at the point singularity 
with $|\Psi| = 0$ due to the vortex cores. 
When the vortex core in one component is filled by the other component, 
the singularity of the first and second term is canceled out, which allows the non-singular 
(coreless) vortices described by the third term. 
The scalar product $[ (\nabla \theta) \times ({\sin} \theta \nabla \phi) ] \cdot d{\bf s}$ 
is equal to the infinitesimal solid angle 
covered by the ${\bf s}$ orientations within the infinitesimal plane $d{\bf s}$. 
By using the pseudospin, the third term can be rewritten as
\begin{eqnarray}
{\bm \omega}_s =  \frac{1}{2} \epsilon_{\alpha \beta \gamma} 
s_\alpha \nabla s_\beta \times \nabla s_\gamma 
\label{vorticityeff2}
\end{eqnarray}
with the Levi-Civita symbol $\epsilon_{\alpha \beta \gamma}$. 
This is known as the Mermin-Ho relation for the ${\bf s}$ texture \cite{MerminHo}.

\subsection{D-brane soliton} 
Next, we turn to the massive case $\Delta g \neq 0$. 
When the condition $\Delta g > 0$ is satisfied, the model has two distinct minima as 
(i) $|\Psi_1| = v$ and $|\Psi_2| = 0$, (ii) $|\Psi_1| = 0$ and $|\Psi_2| = v$. 
In the vacuum (i), the field $|\Psi_2|$ has mass $\sqrt{\Delta g} \, v$. 
The fluctuating field around the vacuum $\delta |\Psi_1| = |\Psi_1| - v$ 
has mass $\sqrt{\lambda} \, v$. In the vacuum (ii), the roles of the $\Psi$ fields 
interchange, as well as their masses. The energies in these two vacua are necessarily 
degenerate because of the ${\bf Z}_2$ symmetry $\Psi_1 \leftrightarrow \Psi_2$ 
apparent in Eq. (\ref{dimlessenergy}), which is spontaneously broken. 
Therefore, there must exist a domain wall interpolating between vacua (i) and (ii). 

\begin{figure}
\begin{center}
\includegraphics[width=0.5 \linewidth,keepaspectratio]{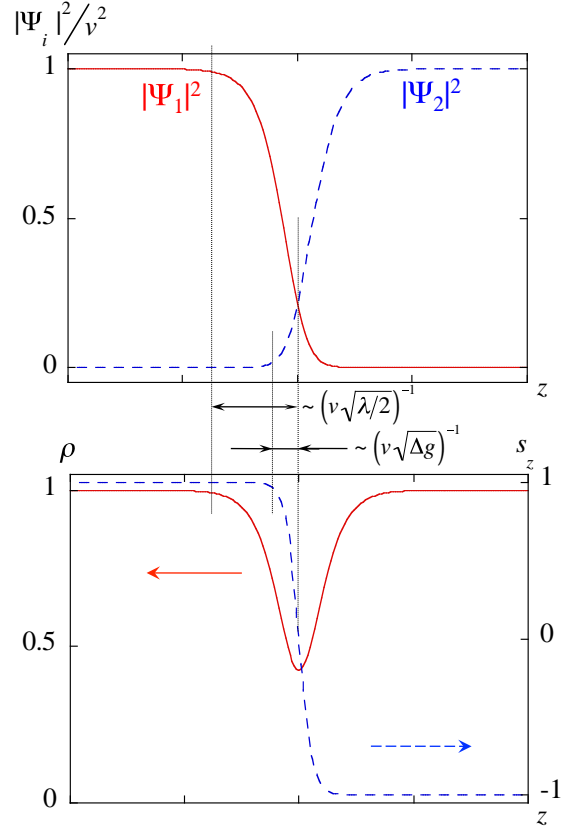}
\caption{Structure of the domain wall in two-component BECs. 
Top panel shows the profile of $|\Psi_1|^2$ and $|\Psi_2|^2$, 
and bottom panel represents the same solution in terms of the total density 
$\rho$ and the pseudospin $s_z$.} 
\label{wallissue}
\end{center}
\end{figure}
Let us assume that the wall lies in the $z=0$ plane and impose the 
following boundary conditions $(\Psi_1, \Psi_2)^{T} \to (v,0)^{T}$ at $z \to - \infty$, 
and $(\Psi_1, \Psi_2)^{T} \to (0,v)^{T}$ at $z \to \infty$. 
The typical profile of the domain wall solution is shown in Fig. \ref{wallissue}.
The amplitude of $\Psi_1$ decreases 
as $v(1-e^{\sqrt{\lambda/2} v z})$ as it approaches to the domain wall, while 
that of $\Psi_2$ increases as $e^{\sqrt{\Delta g} v z}$ from zero moving away from the domain wall. 
Thus the structure of the domain wall has a two-component structure 
with thicknesses $(\sqrt{\lambda} \, v)^{-1}$ and $(\sqrt{\Delta g} \, v)^{-1}$. 
These two length scales are associated with the total density and the pseudospin $s_z$, as 
seen in the bottom panel of Fig. \ref{wallissue}. 
Generally, the solution of the domain wall can be written as 
$\Psi_1=\Psi_{1d} (z-z_0) e^{-i\varphi/2}$ 
and $\Psi_2=\Psi_{2d} (z-z_0) e^{i\varphi/2}$, containing two parameters ---
the wall center $z_0$ and a phase $\varphi = \theta_2 -\theta_1$. 
Both amplitudes of $\Psi_i$ are equal to each other at $z=z_0$. 
The occurrence of $z_0$ is due to the breaking of translational invariance 
by the given wall solution, while $\varphi$ is also due to the breaking of global U(1); 
the relative combination of the phase $\varphi$ is not broken in either 
vacuum (i) or (ii) but is broken along the domain wall, and consequently there 
appears a U(1) Nambu-Goldstone mode localized around the wall. 
These features satisfy a part of the requirement that 
the domain wall in two-component BECs can be referred to as a D-brane 
soliton \cite{KasamatsuD}.

Since the formulation is similar to the strong coupling limit $e \to \infty$ of the Abelian massive 
Higgs model given by Eq. (\ref{multihiggsmassive}), the corresponding analysis in 
Sec. \ref{AbetwomassHigtheory} can be directly applied to our problem. 
According to the Abelian-Higgs model, the vector potential and the adjoint field are not 
independent dynamical valuables, and their forms are given by the profile of $\Psi$ through 
the relations Eq. (\ref{vectorpotphi}) and Eq. (\ref{SigmaMrelation}). 
For $\Delta g > 0$ the potential in Eq. (\ref{dimlessenergy}) can be written as 
\begin{eqnarray}
V(\Psi)=\frac{\lambda}{4} (\Psi^{\dag} \Psi - v^2)^2 
+  \Psi^{\dag} (\Sigma_{\rm eff} {\bf I} - {\bf M})^2 \Psi 
-\delta \mu (|\Psi_1|^2 - |\Psi_2|^2),  \label{potBECmane}
\end{eqnarray} 
where ${\bf M} = {\rm diag.}(M/2, -M/2) $, $M^2 \equiv \rho \Delta g $, and 
\begin{equation}
\Sigma_{\rm eff} = \frac{M}{2} s_z = \frac{M}{2} \frac{|\Psi_1|^2 - |\Psi_2|^2}{|\Psi_1|^2 + |\Psi_2|^2}.
\end{equation} 
The form of Eq. (\ref{potBECmane}) is similar to the potential in Eq. (\ref{multihiggsmassive}), but the mass parameter is 
now dependent on the total density $M=M(\rho)$ \cite{tyuu4}. 
Taking this into account, we can map the hamiltonian Eq. (\ref{dimlessenergy}) 
to the generalized massive O(3) NL$\sigma$M \cite{Kasamatsu2}:
\begin{eqnarray}
H = \int  d^{3} x  \biggl[ \left( \nabla \sqrt{\rho} \right)^2 + \frac{\rho}{4} \sum_{\alpha} 
(\nabla s_{\alpha})^{2} + \rho \left( {\bf v}_s - \tilde{\bf A} \right)^{2}   \nonumber \\
+ \frac{\lambda}{4} (\rho-v^2)^2 + \frac{\Delta g}{4} \rho^2 (1-s_z^2)  -\delta \mu \rho s_z \biggr]. 
\label{nonsigmamodBEC}
\end{eqnarray}
Here, we have additional contributions to Eq. (\ref{massiveNLsm}), namely, the inhomogeneous 
effect of the total density and the kinetic energy of the superflow velocity 
(third term in the right-hand side of Eq.(\ref{nonsigmamodBEC})). 
The latter contribution survives in the GP model 
because the vector potential $\tilde{\bf A}$ is just an external field; 
in the gauge model, $\tilde{\bf A} = {\bf A} = {\bf A}_{\rm eff}$ and this term disappears. 
The density inhomogeneity can be neglected if we consider the limit $\lambda \to \infty$, where 
the radial degree of freedom is frozen as $\Psi^{\dag} \Psi =\rho= v^2$. 
In this limit 
the exact solution of the domain wall as well as the D-brane solitons may be obtained 
from Eq. (\ref{exactsolution}). However, we should pay attention to the properties in this limit, 
because the kinetic energy due to the velocity fields ${\bf v}_s$ becomes divergent if 
there are singular vortices in the system. This divergent contribution may be relaxed 
by depleting the total density at the vortex core. Thus, the fixed total 
density is not a good approximation in the BEC system. 
Detailed analysis shows that, by taking into account the density inhomogeneity, 
the qualitative features of the topological solitons do not change, though 
quantitative profile of the solution is changed through modification of the mass 
parameter $M(\rho) = \sqrt{\rho \Delta g}$ \cite{Kasamatsufull}. 

\begin{figure}
\begin{center}
\includegraphics[width=0.98 \linewidth,keepaspectratio]{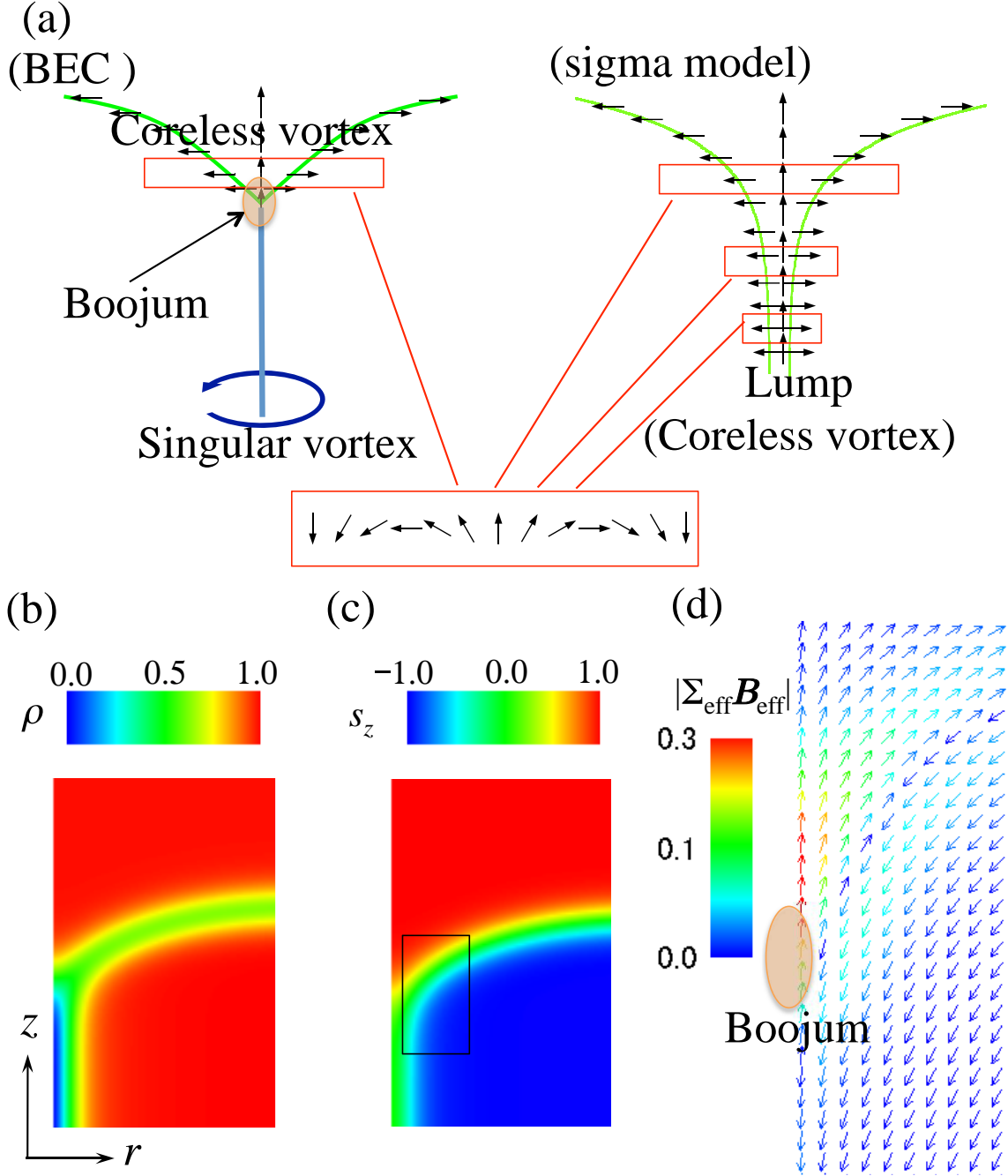}
\caption{Structure of the D-brane soliton in two-component BECs for the simplest situation 
in which a straight vortex is attached to a domain wall. 
(a) Schematic illustration of the D-brane soliton in the NL$\sigma$M 
and the GP model (BEC case). The pseudospin on the domain wall 
has $s_z=0$. The panels (b), (c), and (d) show the numerical solution 
obtained by solving the GP equation with the parameters $\Delta g = 1$ 
and $\delta \mu = 0.01718$. 
(b) Profile of the total density. (c) Profile of $s_z$. 
(d) Vector plot of the boojum flux density
$\Sigma_{\rm eff} {\bf B}_{\rm eff}$ in Eq. (\ref{boochaden}) in the small region 
enclosed by the rectangle in (c).
The vector only shows the radial and $z$-component 
because of the cylindrical symmetric structure. 
The color represents the magnitude 
of the vector $\Sigma_{\rm eff} B_{\rm eff}$. 
The mark represents the region where the boojum charge becomes dense.} 
\label{DbraneBECnume}
\end{center}
\end{figure}
Note that there is a significant difference in the topological structure
of the D-brane soliton between the GP model and NL$\sigma$M.
In the GP model,
the total density $\rho$ decreases exponentially with decreasing $z$ along
the vortex core because the nonrotating component does not enter energetically into the
vortex core [left panel of Fig. \ref{DbraneBECnume}(a)] \cite{tyuu2}.
Since the total density vanishes ($\rho \to 0$) at the singular vortex core, 
the pesudospin is ill-defined there.
In this sense, the lump shrinks to a singular vortex for a finite distance
and thus we can see a boojum connecting a singular vortex to the wall.
This is in contrast to the case in the NL$\sigma$M.
A lump configuration continues to infinity, avoiding the singularity
since $\rho$ is fixed.
Hence, a boojum as a singular defect should be positioned at 
$z \to -\infty$ in this model [right panel of Fig. \ref{DbraneBECnume}(a)].

A typical profile of a boojum structure in two-component BECs is
shown in the bottom of Fig. \ref{DbraneBECnume},
where a single vortex along the $z$-axis in the $\Psi_2$ component is attached
to the domain wall.
The numerical solutions show that the total density $\rho$
remains non-zero along the domain wall
but vanishes along the vortex core for a finite distance.
Therefore, we see a boojum connecting a singular vortex to the wall.
Above the boojum, the vortex becomes a coreless vortex, forming a lump
structure in the $z=$const cross-section. 

This statement can be confirmed by considering the distribution of the boojum charge density 
defined in Eq. (\ref{monochargeden}). Since $t_{m}$ of Eq. (\ref{monochargeden}) 
vanishes for $e \to \infty$, we redefine the boojum 
charge by rescaling as $e^{2} t_m \to t_m$. Using Eq. (\ref{monochargeden}), 
we write the boojum charge of this system as 
\begin{equation}
t_m = \nabla \cdot  (\Sigma_{\rm eff} {\bf B}_{\rm eff} ) = \frac{1}{2} \nabla \cdot  (M s_z {\bf B}_{\rm eff} ). \label{boochaden}
\end{equation}
Using Eqs. (\ref{vectorpotphi}) and (\ref{Phiutrans}) in the NL$\sigma$M limit 
to calculate ${\bf B}_{\rm eff}$, we can show that the boojum 
charge density is identically zero. 
This is because the boojum is relegated to $z = \pm \infty$. 
In contrast, the boojum charge in the BEC is localized around the connecting point of 
the coreless vortex and the singular vortex without moving away to $z= \pm \infty$. 
It is difficult to visualize the boojum charge because it is strongly localized around 
this connecting point, which is needed to calculate the derivative with high accuracy. 
Instead of the boojum charge, in Fig. \ref{DbraneBECnume} (d) we show 
the distribution of the boojum flux density $\Sigma_{\rm eff} {\bf B}_{\rm eff}$. 
The flux is concentrated around the connecting point, which implies that the 
boojum is actually localized there as schematically shown in Fig. \ref{DbraneBECnume} (d).
The distribution of the boojum charge in more general 
configurations of D-brane solitons can be studied by full 3D numerical calculation, 
which will be reported elsewhere. 
 
\section{Conclusion and remarks}\label{concle}
We have shown that an analogue of topological solitons, such as D-brane solitons, 
known in gauge theoretical models can be realized in two-component BECs, 
because the GP model can be regarded as a counterpart of those gauge models. 
The GP model corresponds to the $e \to \infty$ limit of the gauge theory. 
This fact gives the correspondence between two models: (i) The vector potential can 
be represented as a superfluid velocity field. (ii) The adjoint field reduces to the 
$z$-component of the pseudospin density, proportional to the density difference $|\Psi_1|^2 - |\Psi_2|^2$. 
We can obtain the analytic solution of the D-brane solitons in the strongly interacting limit 
$\lambda \to  \infty$, where the radial degree of freedom of $\Psi$ is frozen and the 
model is reduced to the nonlinear $\sigma$ model. 
This approximation can be applied to the BEC problem only in the qualitative level, 
because the density inhomogeneity is crucial around the topological defect. 
Based on the correspondence with the gauge theoretical model, we introduced the boojum charge 
in two-component BECs and showed that the charge is localized around the point 
where a singular vortex terminates on a domain wall.

We believe that atomic BECs are a promising candidate for demonstrating 
an analog of D-brane physics in a laboratory.
The D-brane soliton can be realized as an energetically stable solitonic 
object in phase-separated rotating two-component BECs \cite{KasamatsuD}. 
The long life time of D-brane solitons merits the study of various dynamical phenomena, 
e.g., oscillation modes of strings and branes and nonlinear dynamics such as brane-antibrane annihilation, 
which was proposed as a possible explanation for the inflationary universe in string theory. 
Although brane-antibrane annihilation was demonstrated to show the topological defect creation 
in superfluid $^{3}$He \cite{Bradley}, its physical explanation of the creation mechanism 
of defects is still unclear. In atomic BECs, Anderson {\it et al}. observed the creation 
of vortex rings via the dynamical (snake) instability of a dark soliton \cite{Anderson}, 
where the nodal plane of a dark soliton in one component was filled with the other component and then 
the filling component was selectively removed with a resonant laser beam. 
Here, the snake instability can be understood as a phenomenon similar to tachyon condensation 
in quantum field theory \cite{Takeuchitac}. 
It is quite interesting to note the similarity between tachyon condensation in BECs caused 
by domain-wall annihilation and that by brane annihilation in string theory.  
We also proposed that, when strings are stretched between a brane and an antibrane, namely when 
the filling component has vortices perpendicular to the wall, ``cosmic vortons" can emerge via 
a similar instability \cite{Nittavorton}. 
All of these phenomena can be monitored directly in experiments. 

We do not intend to establish a perfect connection between our D-brane soliton in BECs 
and those in string theory. Strictly speaking, such a perfect connection 
is impossible because the superstring theory is a 10-dimensional theory. 
Also, our system does not contain two important ingredients of original 
string theory: (i) supersymmetry and (ii) relativistic invariance. 
For (i), the authors in Refs. \cite{Gauntlett,Shifman1} have discussed 
an analogue of D-branes in a field theoretical model, which is similar to 
our model. Although their formulation actually possesses supersymmetry, 
it does not give an essential effect to the solution of the D-brane soliton; 
even for a supersymmetric theory, the solutions are 
constructed only by the boson part. In this sense, the supersymmetry 
has no effect on the solution. Regarding the correspondence 
to the original string theory, we note that there are interesting proposals 
\cite{Snoek,Yu:2007xb} in which the supersymmetry 
can be included by preparing the Boson-Fermion mixture in an optical lattice. 
The influences of the optical lattice and fermions in our system 
merit further study.  
For (ii): the relativistic invariance is not significant for our results 
because we confine ourselves to the stationary problem. 
For the relativistic problem, the time evolution of the scalar field 
is governed by the second-order time derivative, while for the non-relativistic case 
the time derivative is of first order. Therefore, the relativistic invariance 
is irrelevant to our stationary problem. Of course, it is important 
when we consider the dynamic behavior of the D-branes or the strings. 
However there have been many studies on non-relativistic D-branes, 
applying them to AdS/condensed matter physics correspondence \cite{correspo}. 

\ack
This work was supported by KAKENHI from JSPS 
(Grant Nos. 21340104, 21740267 and 23740198).
This work was also supported
by the ``Topological Quantum Phenomena'' 
(Nos. 22103003 and 23103515)
Grant-in Aid for Scientific Research on Innovative Areas 
from the Ministry of Education, Culture, Sports, Science and Technology 
(MEXT) of Japan.


\begin{thebibliography}{99}
\bibitem{tyuu3}
The system is invariant under the global SU(2) transformation and it can be 
chosen as $\Phi=(\Phi_1,0)$ without loss of generality. ${\rm U(1)}_{\rm L+G}$ means 
that the vacuum is invariant under the combined transformation 
$e^{i \alpha} e^{-i \sigma_z \alpha} (\Phi_1,0) = (\Phi_1,0)$.
\bibitem{tyuu4}
Actually, we have the choice of including $\sqrt{\rho}$ in $\Sigma_{\rm eft}$ and $M$ or not. 
We include this factor because (i) the potential form of Eq. (\ref{potBECmane}) 
can imitate that of Eq. (\ref{multihiggsmassive}), (ii) the inclusion can avoid the fact that 
the boojum flux density $\Sigma {\bf B}_{\rm eff}$ 
in Eq. (\ref{boochaden}) is ill-defined at the singular vortex core at which $\rho=0$.
\bibitem{tyuu2}
This argument is correct when $\delta \mu$ is relatively large.
The nonrotating component can invade into the core for smaller $\delta \mu$ \cite{Kasamatsufull}.
\end{thebibliography}
\end{document}